\documentclass[conference]{IEEEtran}
\usepackage{scalerel}
\usepackage{tikz}
\usetikzlibrary{svg.path}

\definecolor{orcidlogocol}{HTML}{A6CE39}
\tikzset{
  orcidlogo/.pic={
    \fill[orcidlogocol] svg{M256,128c0,70.7-57.3,128-128,128C57.3,256,0,198.7,0,128C0,57.3,57.3,0,128,0C198.7,0,256,57.3,256,128z};
    \fill[white] svg{M86.3,186.2H70.9V79.1h15.4v48.4V186.2z}
                 svg{M108.9,79.1h41.6c39.6,0,57,28.3,57,53.6c0,27.5-21.5,53.6-56.8,53.6h-41.8V79.1z M124.3,172.4h24.5c34.9,0,42.9-26.5,42.9-39.7c0-21.5-13.7-39.7-43.7-39.7h-23.7V172.4z}
                 svg{M88.7,56.8c0,5.5-4.5,10.1-10.1,10.1c-5.6,0-10.1-4.6-10.1-10.1c0-5.6,4.5-10.1,10.1-10.1C84.2,46.7,88.7,51.3,88.7,56.8z};
  }
}

\newcommand\orcidicon[1]{\href{https://orcid.org/#1}{\mbox{\scalerel*{
\begin{tikzpicture}[yscale=-1,transform shape]
\pic{orcidlogo};
\end{tikzpicture}
}{|}}}}
\usepackage{cite}
\usepackage{hyperref}
\usepackage{tabularx}

\ifCLASSINFOpdf

\else
 
\fi

\usepackage{graphicx}
\begin{document}
%
\title{Highly-Multiplexed Superconducting Detector Readout: Approachable High-Speed FPGA Design}

\author{\IEEEauthorblockN{Jennifer Pearl Smith\IEEEauthorrefmark{1}  \orcidicon{0000-0002-0849-5867},
John I. Bailey, III.\IEEEauthorrefmark{2} \orcidicon{0000-0002-4272-263X},  and
and Benjamin A. Mazin\IEEEauthorrefmark{3}  \orcidicon{0000-0003-0526-1114}}
\IEEEauthorblockA{Department of Physics and Astronomy,
University of California, Santa Barbara\\
Santa Barbara, CA, 93106, USA\\
Email: \IEEEauthorrefmark{1}jennifer\_smith@ucsb.edu,
\IEEEauthorrefmark{2}baileyji@ucsb.edu,
\IEEEauthorrefmark{3}bmazin@ucsb.edu}}
\maketitle

\begin{abstract}
This work presents the design and preliminary performance of a highly-multiplexed superconducting detector readout. The readout system is implemented on the Xilinx ZCU111 RFSoC Evaluation Board. The current design uses 12\% of the DSPs, 60\% of the LUTs, 20\% of the FFs, and 30\% of the BRAM and makes timing at 512 MHz. The system uses two integrated ADCs and DACs running at 4.096 GSPS to read out 2,048 superconducting detectors. This work targets a 2x increase in the number of superconducting detectors processed per board with 80\% less power than previous readout schemes. The open-source design leverages modern FPGA productivity tools including Vivado High-Level Synthesis to create all custom IP blocks, PYNQ to test and verify individual IP and develop Python drivers, and Vivado ML Intelligent Design Runs to close timing. We emphasize strategies for achieving timing closure without custom HDL which we expect to be useful for superconducting device groups looking to utilize FPGAs in high-performance applications without specialized knowledge in FPGA design.
 
\end{abstract}


%
\IEEEpeerreviewmaketitle

\section{Introduction}
Superconducting devices represent a class of breakthrough technologies enabling detection and manipulation of quantum signals \cite{arute_quantum_2019, amaral_constraints_2020, de_visser_phonon-trapping-enhanced_2021}. Researchers are working to build large arrays of superconducting devices to support quantum error correction for quantum computing and mega-pixel-scale scientific cameras for single-photon-sensing applications \cite{stassi_scalable_2020, truitt_development_2020, walter_mkid_2020}. These large arrays impose a technical challenge of reading out thousands of highly-multiplexed device signals in real time.

FPGAs offer highly-customizable, parallel computing architectures, large I/O bandwidths, and low power consumption, enabling rapid, real-time processing of highly-multiplexed superconducting array signals. Researchers have developed multiple custom, FPGA-based superconducting array readouts; however, current systems still rely on hardware and toolflows over a decade old\cite{ryan_hardware_2017, kernasovskiy_slac_2018, fruitwala_second_2020, gordon_open_2016, madden_development_2017}. Previous projects have stagnated in part due to the complexity involved in custom FPGA design which makes it tedious to upgrade and maintain systems. As a result, systems have failed to keep pace with modern hardware and are now excessively bulky and power-intensive which limits system scale and complicates deployment.

We seek to improve state-of-the-art superconducting device readout by migrating to a RFSoC platform\footnote{Xilinx ZCU111 Evaluation Board}. The RFSoC provides high-speed, integrated analog data converters, enabling an 80\% decrease in the weight, volume, and power of the readout electronics\cite{RFSoCWP}. Several groups are developing RFSoC-based superconducting array readouts\cite{stefanazzi_qick_2021, sinclair_development_2020, gebauer_state_2020}, but our approach is unique in targeting a more aggressive 512 MHz FPGA fabric speed while simultaneously moving away from custom HDL implementation and vendor-entrenched verification. 

In this paper we present our development of a readout for superconducting Microwave Kinetic Inductance Detectors (MKIDs) with findings applicable to a broad range of superconducting device readouts . Our system targets a 2x increase in the number of detectors read out while using 80\% less power than our previous MKID readout system\cite{fruitwala_second_2020}. We uniquely leverage modern FPGA productivity tools including Vitis High-Level Synthesis (HLS)\cite{noauthor_vitis_2021}, Vivado ML Intelligent Design Runs (IDR)\cite{noauthor_elevating_2021}, and Python Productivity for ZYNQ (PYNQ)\cite{noauthor_pynq_nodate} to create a modern design flow enabling rapid development without sacrificing performance. Developing logic blocks in HLS allows us to write C/C++ code instead of HDL, leading to faster workflows with more readable, approachable, and portable code. IDR provides intelligent synthesis and implementation strategies which yields increased quality of results and facilitates timing closure at 512 MHz despite using HLS-generated code. PYNQ provides simplified testing with multiple Overlay designs and \textit{in-situ} development of easy-to-use Python drivers. 
This combination of modern tools and techniques prepares our system for simplified upgrading to future devices and speed-grades and maintenance by non HDL-equipped researchers and staff. We discuss the design and implementation strategies and report on our progress. Special care is given to lessons learned regarding timing closure techniques so that other groups may benefit. The current state of the project is available on \href{https://github.com/MazinLab/MKIDGen3}{Github}\footnote{\href{https://github.com/MazinLab/MKIDGen3}{https://github.com/MazinLab/MKIDGen3}}.

\section{System Design Overview}
\begin{figure}
\includegraphics[width=3.5in]{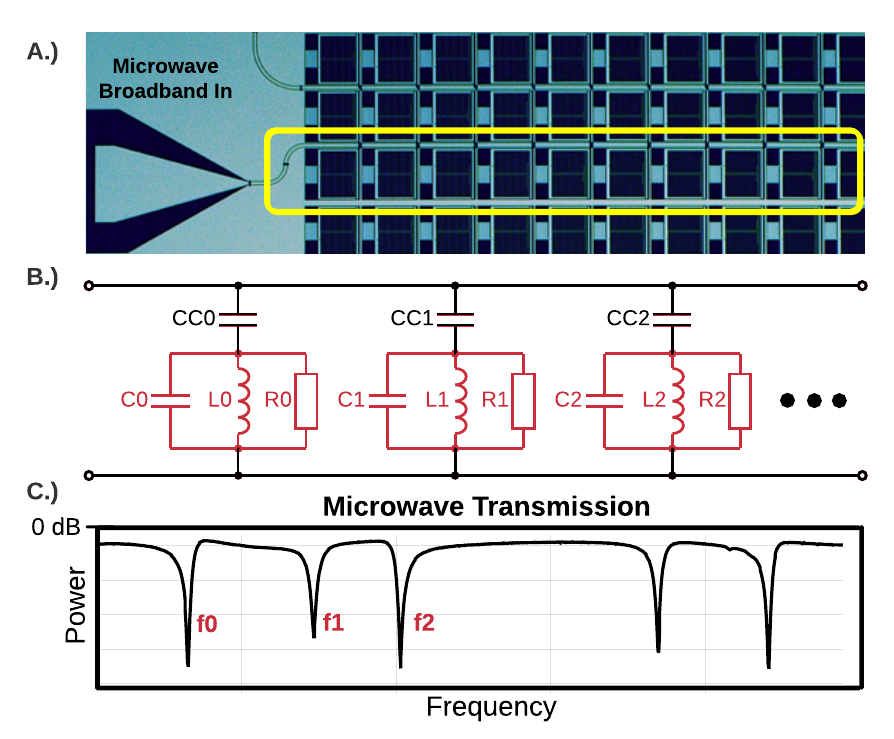}
\caption{A.) Optical microscope image of a MKID array. Broadband microwave signals are supplied to the chip via the triangular feedline on the left. Nine and a half individual MKID detectors, each composed of a superconducting inductor and capacitor, are highlighted in the yellow rectangle. B.) Circuit schematic representation of MKID devices multiplexed on a single feedline. Each MKID detector (red) has a unique resonant frequency in the 4-8 GHz range. Up to 2,048 MKIDs are connected to a single microwave feedline through coupling capacitors (black). C.) Representative microwave transmission of a single microwave feedline. Dips in the power trace out the resonate frequencies corresponding to each MKID. The resonators are spaced approximately every 2 MHz but actual resonant frequencies vary considerable between MKID chips.} \label{fig:multiplex}
\end{figure}

MKID detectors are lithographed superconducting LC resonators, each with a unique resonance between 4-8 GHz \cite{Day2003}. Up to 2,048 MKIDs are coupled to a single microwave feedline with resonances spaced $\sim$2 MHz apart\cite{Szypryt2017Large-formatAstronomy}. We use frequency-division-multiplexing to simultaneously drive and monitor every MKID resonator at its resonant frequency \cite{fruitwala_second_2020}. An image of an MKID chip along with two representations of MKID multiplexing strategy is shown in Fig.\ref{fig:multiplex}.
 
\begin{figure}
\includegraphics[width=3.5in]{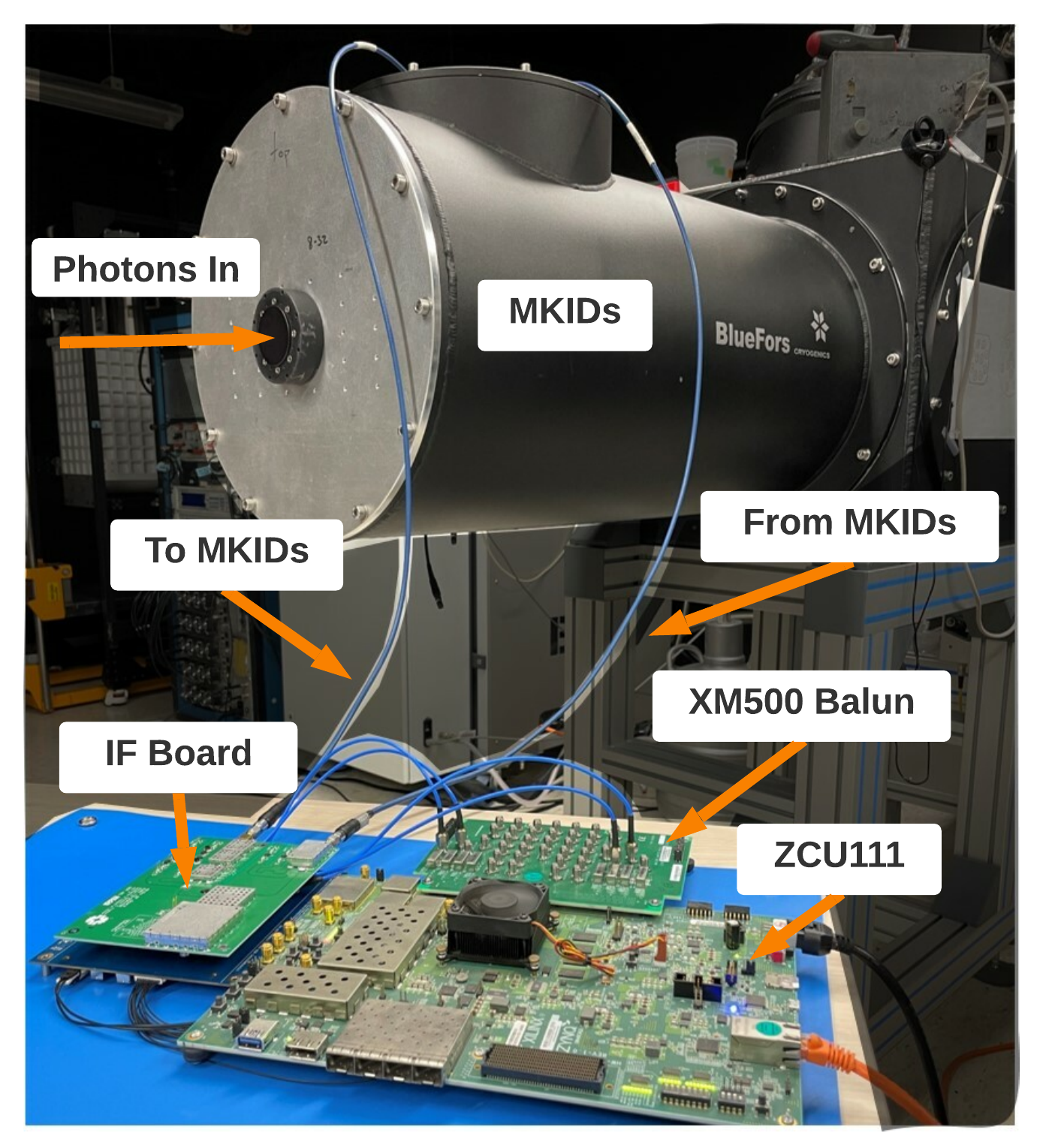}
\caption{Picture of planned MKID readout system hardware. The MKID detectors are housed in a cryogenic refrigerator (black cylinder). An IF board modified from \cite{fruitwala_second_2020} translates between the 4-8 GHz RF band and $\pm$2 GHz IF band. The IF board (green, left) is powered by a custom carrier board (blue) and will be controlled by the RFSoC CPU over USB or GPIO. The IF board is connected to the ZCU111 via the attached XM500 Balun card.}\label{fig:readout_hardware}
\end{figure}

\subsection{Intermediate Frequency Conversion}
In order to access the 4-8 GHz MKID frequencies, an intermediate frequency (IF) board is used to convert 4 GSPS, 2 GHz in-phase (I) and quadrature (Q) IF signals to and from the 4-8 GHz RF band. The IF system interfaces with the Xilinx XM500 balun card which attaches to the ZCU111 on the IF side and the MKID device lines on the RF side. The IF board is a modified version of the custom IF board described in \cite{fruitwala_second_2020}. An image of the   readout system hardware in a testing configuration is shown in Fig. \ref{fig:readout_hardware}.

\subsection{Setup and Calibration}
Before the MKIDs can be read out, the RFSoC system must characterize the detectors. This is accomplished by sending broadband 4-8 GHz signals to the device to produce data similar to that shown in Fig. \ref{fig:multiplex}C. The data is sent off the RFSoC device to a control computer which runs a deep learning pipeline capable of identifying and cataloging resonance dips similar to \cite{fruitwala_end--end_2021}. During the calibration phase, each detector's response to photon events is recorded and used to generate a custom matched filter using the technique established in  \cite{moseley_advances_1988,alpert_note_2013} to  enabling accurate photon characterization.

\subsection{Data Collection}
\begin{figure}
\includegraphics[width=3.5in]{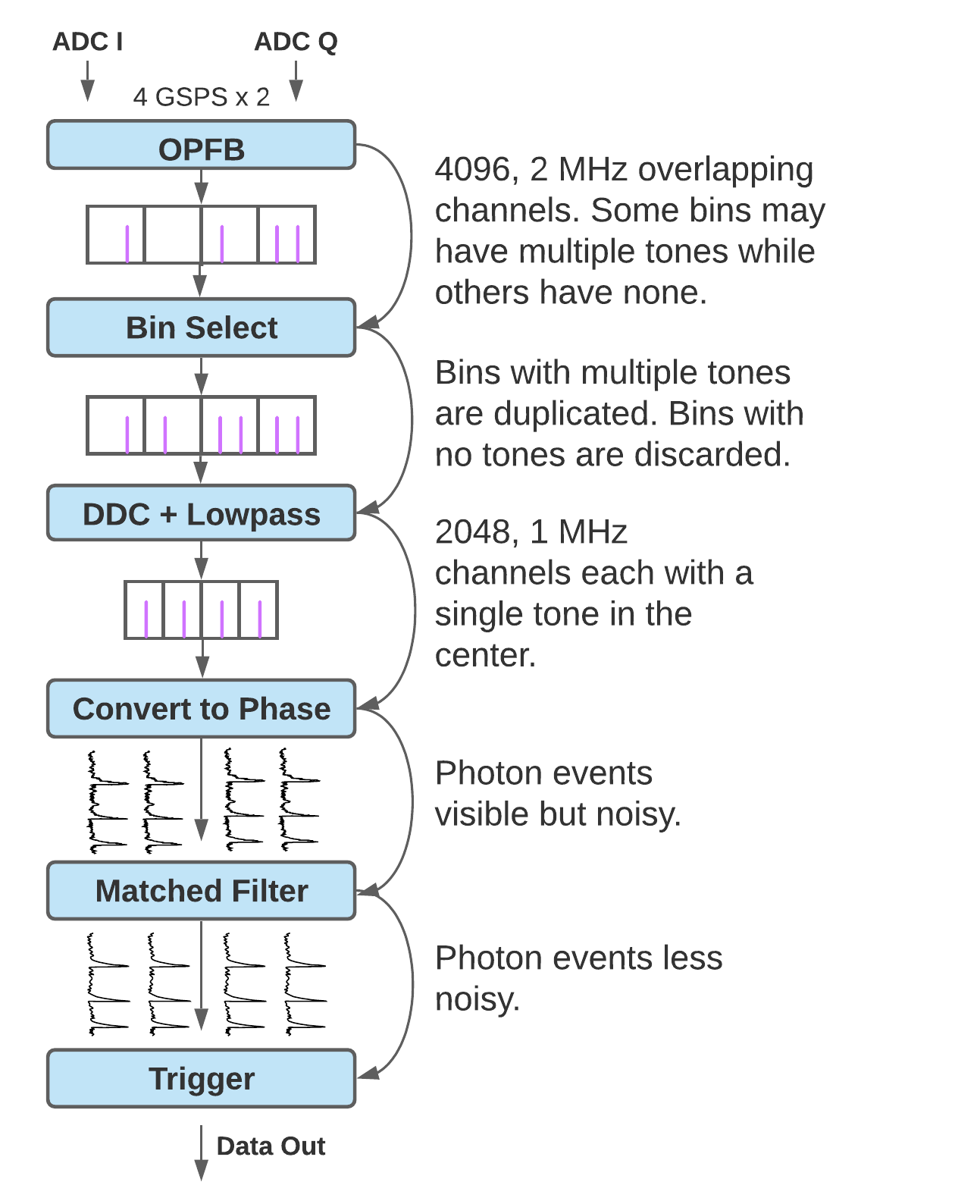}
\caption{Overview of the key digital signal processing steps for multiplexed MKID readout. The programmable logic accepts dual 4.096 GSPS input streams and processes the data in real time to record and characterize individual photons hitting the MKID array.}\label{fig:dsp}
\end{figure}

Once characterized, Python software generates a waveform comprised of the MKID resonant frequencies for continuous replay by two RFSoC DACs. Two RFSoC ADCs continuously sample the return waveform and FPGA logic channelizes each resonant frequency, converts to phase, and applies the matched filter. The resulting phase time streams are monitored for photon events which are then recorded as a photon time-series for later analysis and image/video reconstruction. The signal processing steps along with their intermediate data products are shown in Fig. \ref{fig:dsp}.

\begin{figure*}
\includegraphics[width=7in]{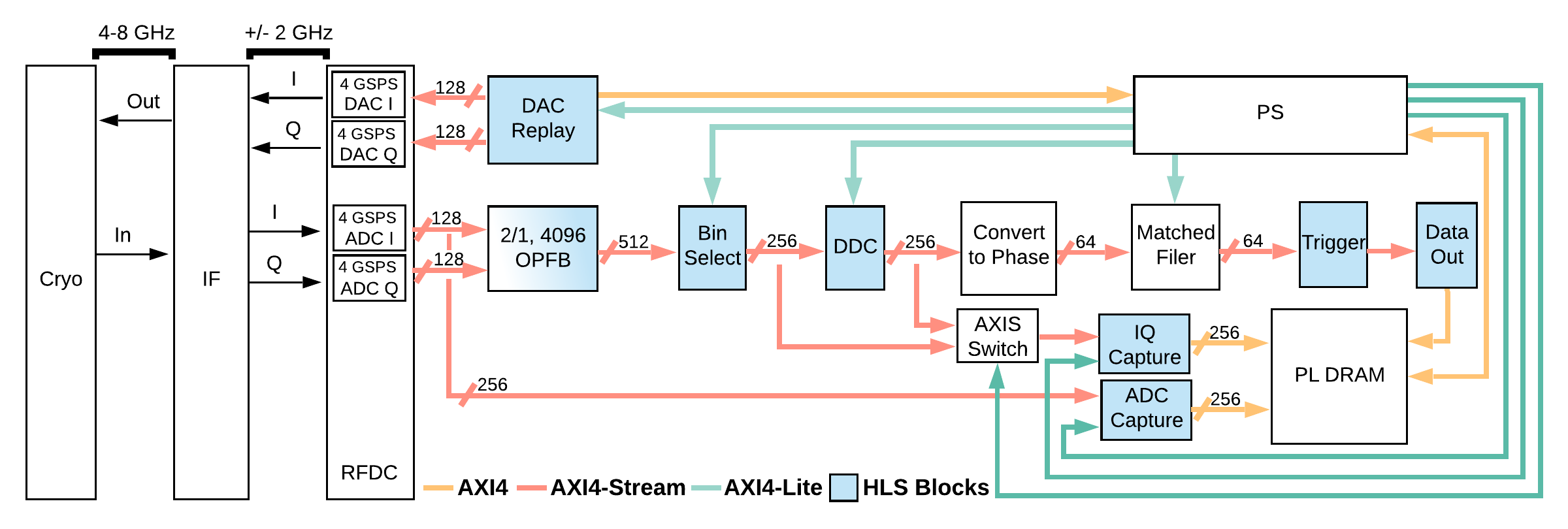}
\caption{High-level block design detailing the main IP blocks and the data routing. HLS blocks are shown in blue while existing Xilinx IP are shown in white. The OPFB block is shaded blue and white because it incorporates HLS for internal stream routing but uses existing Xilinx IP for mathematical operations, see \cite{smith_high-throughput_2021} for details. The main signal processing pathway, including the OPFB through the Data Out block and the capture cores, operates on a 512 MHz clock domain. The DAC Replay operates on a 256 MHz clock domain. The Memory Interface Generator (PL DRAM) operates on a 300 MHz clock by default.} \label{fig:BlockDesign}
\end{figure*}
\section{Implementation}
The design is built using Vivado HLx 2021.2 with an IP-Integrator flow. An overview of the IP regions and data flow is shown in Fig.\ref{fig:BlockDesign}. After being sampled by the 4.096 GSPS I and Q ADCs, the data is separated into coarse channels using a 4096-channel, 2/1 oversampled polyphase filter bank modified from \cite{smith_high-throughput_2021}. Overlapping channels are used to guarantee any MKID frequency can be channelized with minimal attenuation. An HLS bin-selection block duplicates channels with more than one tone and drops channels with no tones, leaving 2048, 2 MHz bins. These channels are further refined by an HLS direct digital down-conversion (DDC) block which uses a direct-digital-synthesis technique to migrate the tones to the bin centers. A 500 KHz lowpass filter is applied and the result is down-sampled by 2, yielding 2048, 1 MHz wide bins with DC tones. A suite of cordic blocks implementing arc-tangent convert the complex I and Q samples to phase. We use reloadable Xilinx FIR cores to apply the custom matched-filter coefficients determined during setup to each phase time stream. A trigger unit monitors the phase time streams and records photon events. 

We use the PYNQ framework to significantly simplify interacting with the programmable logic cores and facilitate rapid test bench and example development through its Jupyter notebook server hosted on the processing system (PS). We have used PYNQ to develop Python drivers to control HLS IP over AXI4-Lite and allocate PL DRAM for high-speed buffering. We use the ADC Capture and IQ Capture blocks to trigger capture events from various places in the signal processing pipeline to the PL DRAM. The PS is then able to read data from the PL DRAM, carrying out both further processing and facilitating off device analysis including key setup and calibration steps.

 The main signal processing data path operates on a 512 MHz clock domain created by a clocking wizard MMCM and sourced from the RFDC ADC tile PLL. The DAC Replay operates on a 256 MHz clock domain sourced from the DAC tile PLL. The PS supplies a 100 MHz clock for low-performance command and control via AXI4-Lite. The Memory Interface Generator (PL DRAM) operates on a 300 MHz clock by default.

\section{Timing Considerations}

This design features an aggressive 512 MHz clock rate in the DSP fabric despite using HLS-generated IP. In pursuing this goal, we have 
developed several recommendations for enhancing timing performance both for HLS specific projects and the design as a whole. We share our thoughts below. 

We first improved timing results by freeing logic and routing resources where possible, giving the tools the flexibility to find a solution. Our design features a 2 MiB waveform look-up-table that default implemented as BRAM. We switched this block to use URAM, which gave the resource-intensive FIR filters more flexibility to trade LUTRAM and BRAM as needed which improved timing performance. We also benefited from eliminating resets wherever possible. We grounded every reset in the design except those required by the Memory Interface generator (PL DRAM) and RF Data Converter. 

We also closely examined the use of AXI4 Interconnects in the design. At times we found Vivado instantiated a full AXI4 bus when only AXI4-Lite was needed. We corrected this by enforcing AXI4-Lite via a Protocol Conversion block. We also found instantiating separate AXI4 Interconnects for each clock domain reduced the number of internal clock domain crossings and enhanced timing performance. 

HLS supports the use of a separate AXI4-Lite clock for control signals. Counterintuitively, we found that using two clocks in one HLS IP intensified logic resources and hurt timing performance. We recommend clocking the AXI4-Lite control signals at the block data rate.

Lastly, we recommend the use of Vivado ML Intelligent Design runs and QoR suggestions to help close timing. These features enable discovery and programmatic application of optimizations less experienced FPGA developers might not be aware of. We found the suggestions were particularly useful for fixing sub-optimal synthesis in HLS blocks.

\section{Current Status}

We have implemented all IP systems shown in Fig. \ref{fig:BlockDesign}. The OPFB, DAC Replay, Bin Select, DDC and ADC capture blocks have all been verified on the hardware in isolation using a suite of PYNQ Overlays to generate test data and visualize and verify output data products. We have implemented the full system as shown in Fig. \ref{fig:BlockDesign}, except for the trigger and data out blocks, on the FPGA to get an accurate estimate of resources and ensure we will be able to close timing.

\begin{figure}
\includegraphics[width=3.5in]{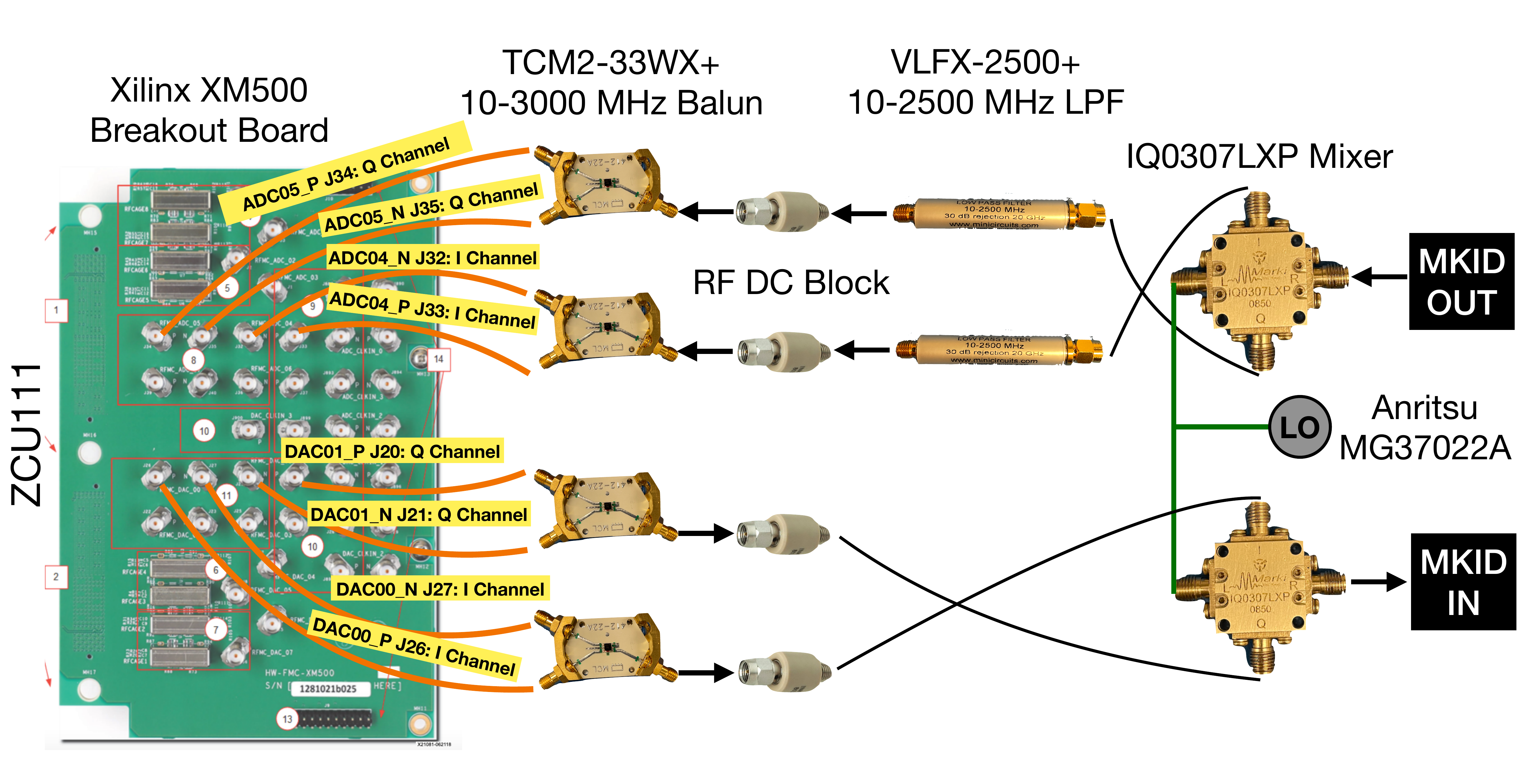}
\caption{Schematic visualization of the test setup. Bench-top IQ mixers and a programmable LO are used to convert the $\pm$ 2 GHz IF band to and from the 4-8 GHz RF band required by the MKID array. Anti-aliasing filters are present only on the ADC input.}\label{fig:test_schematic}
\end{figure}

We have successfully used the DAC replay block in conjunction with the RFDC and ADC Capture to drive and sample a superconducting detector. The IF board is still undergoing testing so we instead used bench-top Marki IQ0307LXP IQ mixers and an Anritsu MG37022A LO to mix a 500 MHz DAC output to and from the 4-8 GHz superconducting detector band. The hardware setup for this test is shown in Fig. \ref{fig:test_schematic}. We captured ADC data to the PL DRAM and visualized it using the Jupyter notebook running on the RFSoC embedded CPU. We can clearly see the 500 MHz tone, the image tone, and the LO signal. Higher order harmonics are also present likely because we did not have anti-aliasing filters for the DAC output. We expect a cleaner spectrum once we integrate the IF board for the up-conversion as it provides more sophisticated filtering and programmable attenuation. This experiment confirms the DAC Replay, RFDC, and ADC Capture are working as expected in the integrated system.

Currently, we are continuing to integrate signal processing blocks in the full system as shown in Fig. \ref{fig:BlockDesign} and are developing python drivers to simplify operation. We are also working to integrate the IF board into the hardware setup. We plan to perform similar verification tests with the data processing chain and IQ capture core.

\begin{figure}
\includegraphics[width=3.5in]{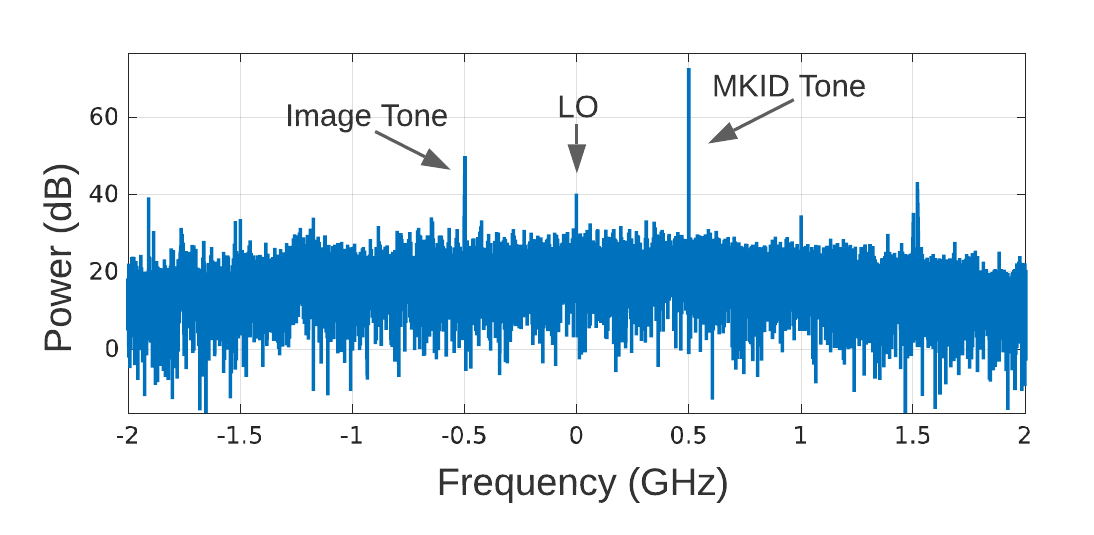}
\caption{Full, 4 GHz spectrum captured by two RFSoC ADCs running at 4.096 GSPS. The input 500 MHz tone is visible along with the image tone and the LO (center). Harmonics and harmonic images are also present due to the lack of anti-aliasing filters on the DAC output.}\label{fig:mkid}
\end{figure}

\section{Conclusion}
We have demonstrated the utility and success of our modern approach to high-speed FPGA design for superconducting array readouts. Our utilization of HLS, IDR, and PYNQ enhance the usability and portability of our system all while achieving high-throughput performance. The resulting system can be adapted with high-level changes to C/C++ code and readily tested on hardware with an intuitive Jupyter Notebook interface. As a result of this technique, we are well-positioned to take advantage of future device innovations as they become commercially available and our system is easier to maintain for a broader audience. Our target RFSoC hardware provides a dramatic reduction in the weight, volume, and power of the readout electronics which enables large-scale detector arrays and deployment in remote environments. All control software, custom IP, and overlay designs are open source and available on  \href{https://github.com/MazinLab/MKIDGen3}{Github}\footnote{\href{https://github.com/MazinLab/MKIDGen3}{https://github.com/MazinLab/MKIDGen3}}.


\section*{Acknowledgment}

The authors would like to thank the Xilinx High-Speed Timing Closure working group for their valuable insights on using IDR and QoR strategies to help our design close timing.

J. P. Smith is supported by a NASA Space Technology Research Fellowship under grant number 80NSSC19K1126.


\bibliographystyle{IEEEtran}
\bibliography{references.bib}

\end{document}